\documentclass[11pt]{article}
\usepackage{moriond,epsfig}

\def\be{\begin{equation}}
\def\ee{\end{equation}}
\def\bea{\begin{eqnarray}}
\def\eea{\end{eqnarray}}

\begin{document}
\vspace*{4cm}
\title{HERA PRECISION MEASUREMENTS AND IMPACT FOR LHC PREDICTIONS}

\author{ V. Radescu  on behalf of the H1 and ZEUS Collaborations}

\address{Heidelberg Physikalisches Institut, Philosophenweg 12, D-69120 Heidelberg, Germany}

\maketitle\abstracts{
A QCD fit analysis to the combined HERA inclusive deep inelastic cross sections measured by the H1 and ZEUS collaborations for $e^\pm p$  scattering to extract HERAPDF sets is presented. The results are used for predictions of $ p\bar{p}$ processes at Tevatron and 
$pp$ processes at the LHC.
The QCD analysis has been extended to include the combined HERA II measurements at high $Q^2$
resulting in the HERAPDF1.5 sets, with full estimation of uncertainties. 
The precision of the new PDFs at high $x$ is considerably improved, particularly in the valence sector.
In addition, inclusion of the HERA jet data allows for a precise determination of the strong coupling. 
Moreover, inclusion of  the preliminary combined HERA charm data provides constraints for the optimal value of the charm mass used in QCD theory models which may account for some of the differences among global PDF fits.
}
\section{Combined H1 and ZEUS Cross Section Measurements}
The main information on proton structure functions comes from the Deep Inelastic Scattering (DIS) collider experiments H1 and ZEUS at HERA. 
 Measurements at HERA go well beyond the phase space accessible by fixed target experiments  with an extended kinematic range of  $0.045 < Q^2 < 30\,000$~GeV$^2$ and $6\times10^{-5}<x<0.65$.
 To further benefit from the precision of the measurements the H1 and ZEUS cross sections  are combined, resulting in the most consistent and precise DIS inclusive double differential cross-section measurement of neutral and charged current $e^\pm p$ scattering  to date \cite{herapdf}. The
data combination "procedure" has been repeated to include as well the preliminary high precision measurements at high $Q^2$ from the second run period of HERA.
Therefore, these
accurate measurements can be used as the sole input to the QCD analysis to determine the proton parton distribution functions
(PDFs) as described in the following sections, which can be used then for precise predictions for LHC processes.
\section{QCD Analysis settings}\label{sec:analysis}
HERA PDFs are determined from QCD fits to HERA data alone.
Only  the region where perturbative QCD is valid, data above $Q^2_{\rm{min}}=3.5$~GeV$^2$ are used in the central fit. The HERA data have a minimum invariant mass of the hadronic system, $W$, of $15$ GeV, such that they are in a kinematic region where there is no sensitivity to non-perturbative effects common to fixed target data.

The fit procedure starts by parametrising PDFs at a starting scale  $Q^2_0=1.9~ \rm GeV^2$,
chosen to be below the charm mass threshold.
The QCD settings are optimised for HERA measurements of proton structure functions
which are dominated by gamma exchange, therefore the 
    parametrised PDFs are the valence distributions
 $xu_v$ and  $xd_v$,  the gluon distribution $xg$, and the $u$-type and $d$-type sea quark distributions
$x\bar{U}$, $x\bar{D}$, where $x\bar{U} = x\bar{u}$, 
$x\bar{D} = x\bar{d} +x\bar{s}$. Using a simple parametrisation form with the normalisation parameters constrained by  
the QCD sum-rules and relying on additional assumptions \cite{herapdf} the number of free paramaters reduces to $10$ for the fits to HERA I data.
However, the more data become available, the more constraining assumptions can be released, therefore the number of free parameters once HERA II data are added is increased, allowing for more freedom to the gluon and valence distributions, HERAPDF1.5f. For example addition of the preliminary combined charm data, or jet data from H1 \cite{jetsh11,jetsh12} and ZEUS \cite{jetsz1,jetsz2} allows for more flexibility of the gluon distribution.  
The PDFs are then evolved using the DGLAP evolution equations \cite{qcdnum}  at NLO  and NNLO in the $\overline{MS}$ scheme with the
renormalisation and factorisation scales set to $Q^2$.
The QCD predictions for the structure functions 
are obtained by convoluting 
the PDFs with the calculable coefficient functions taking into account mass effects for the heavy quarks based on the general mass variable flavour scheme \cite{tr}. 
The PDF uncertainties at HERA are classified in three categories: experimental, model, and parametrisation.
The consistency of the input data set 
and its small systematic uncertainties enables us 
to calculate the experimental uncertainties on the PDFs using the 
$\chi^2$ tolerance $\Delta\chi^2=1$. 
The model uncertainties are evaluated by varying the input assumptions, as performed in \cite{herapdf}.
The parametrisation uncertainty  
is estimated as an envelope which is formed as a maximal deviation at each $x$ value from the central fit.
\section{Results and Comparisons}\label{sec:results}

\begin{figure}[h!]
\centering
\includegraphics[width=0.495\textwidth]{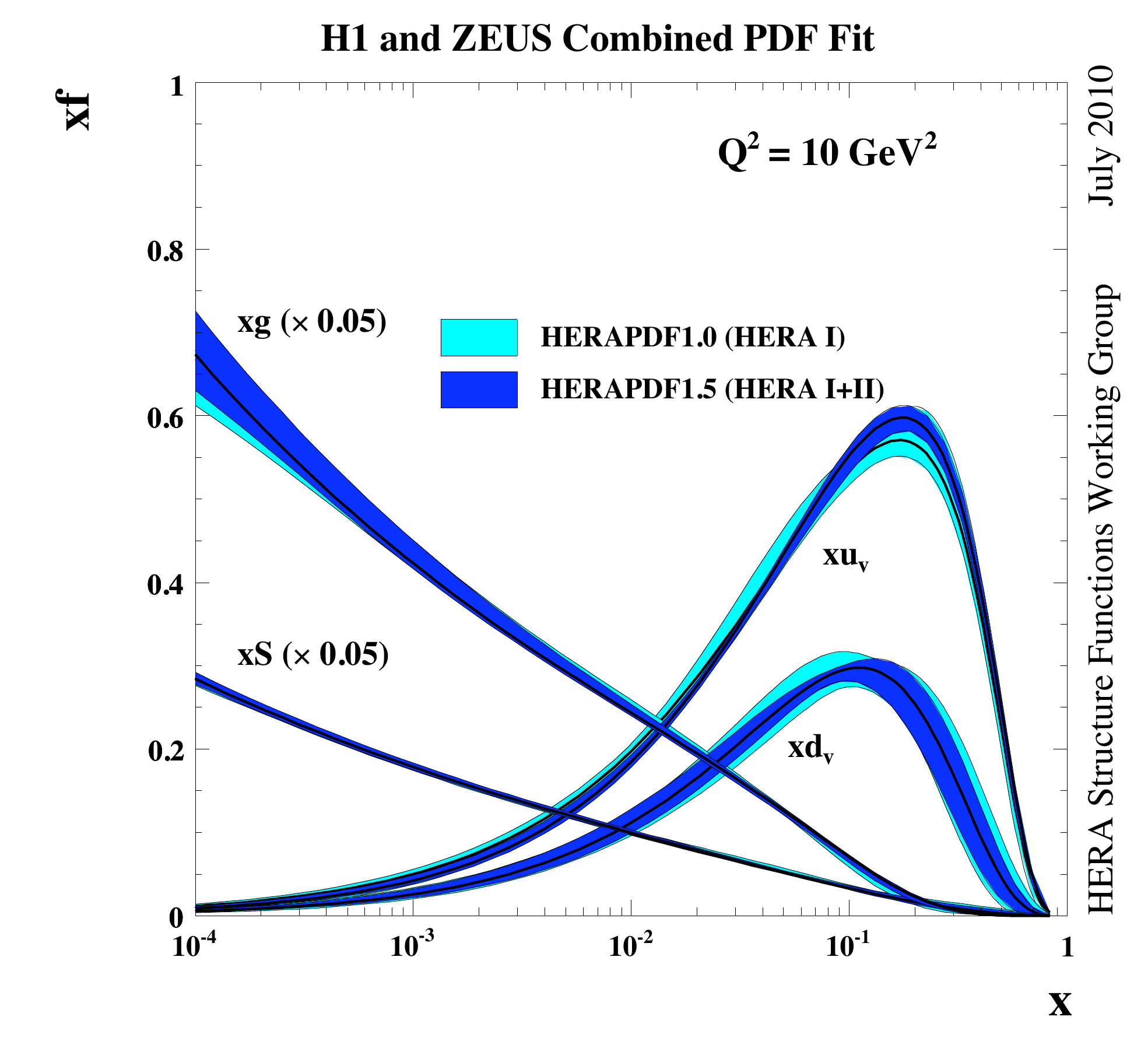}
\includegraphics[width=0.495\textwidth]{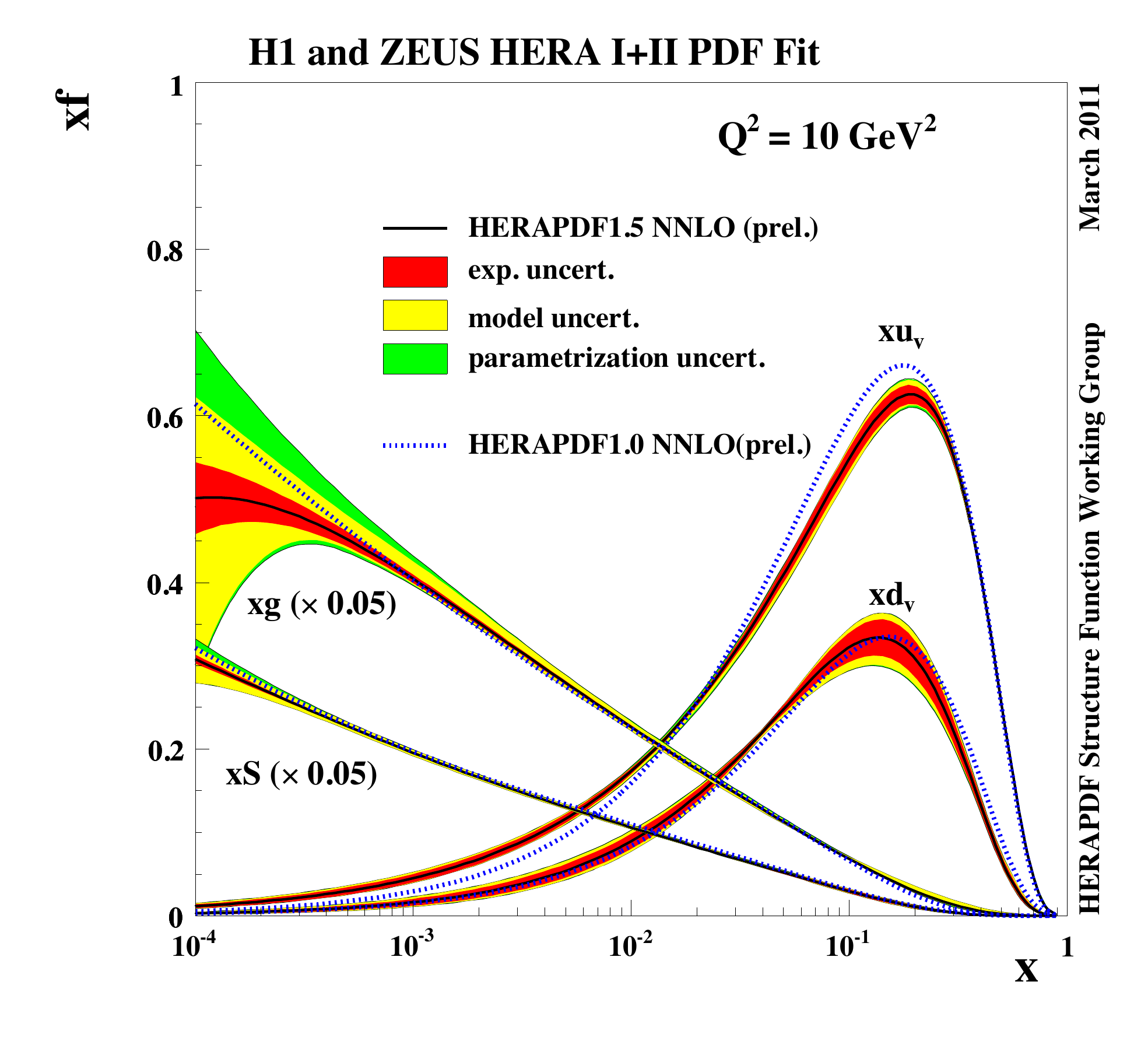}
\caption{On the left hand side:
the comparison between HERAPDF1.0  (light colour)  based on the HERA I data  and HERAPDF1.5 (dark colour) based on the HERA I and  II data, using the total uncertainty band at $Q^2=10$ GeV$^2$ with gluon, sea  (which are scaled by a factor of 0.05) and valence distributions.
On the right hand side: the summary plot for HERAPDF1.5 at $Q^2=10$ GeV$^2$ at NNLO with the uncertainties including
the experimental (red), model (yellow) and the PDF parametrisation (green), compared to the central fit for HERAPDF1.0 at NNLO.}
 \label{Fig:1}
\end{figure}

\begin{figure}[h!]
\centering
\includegraphics[width=0.4\textwidth]{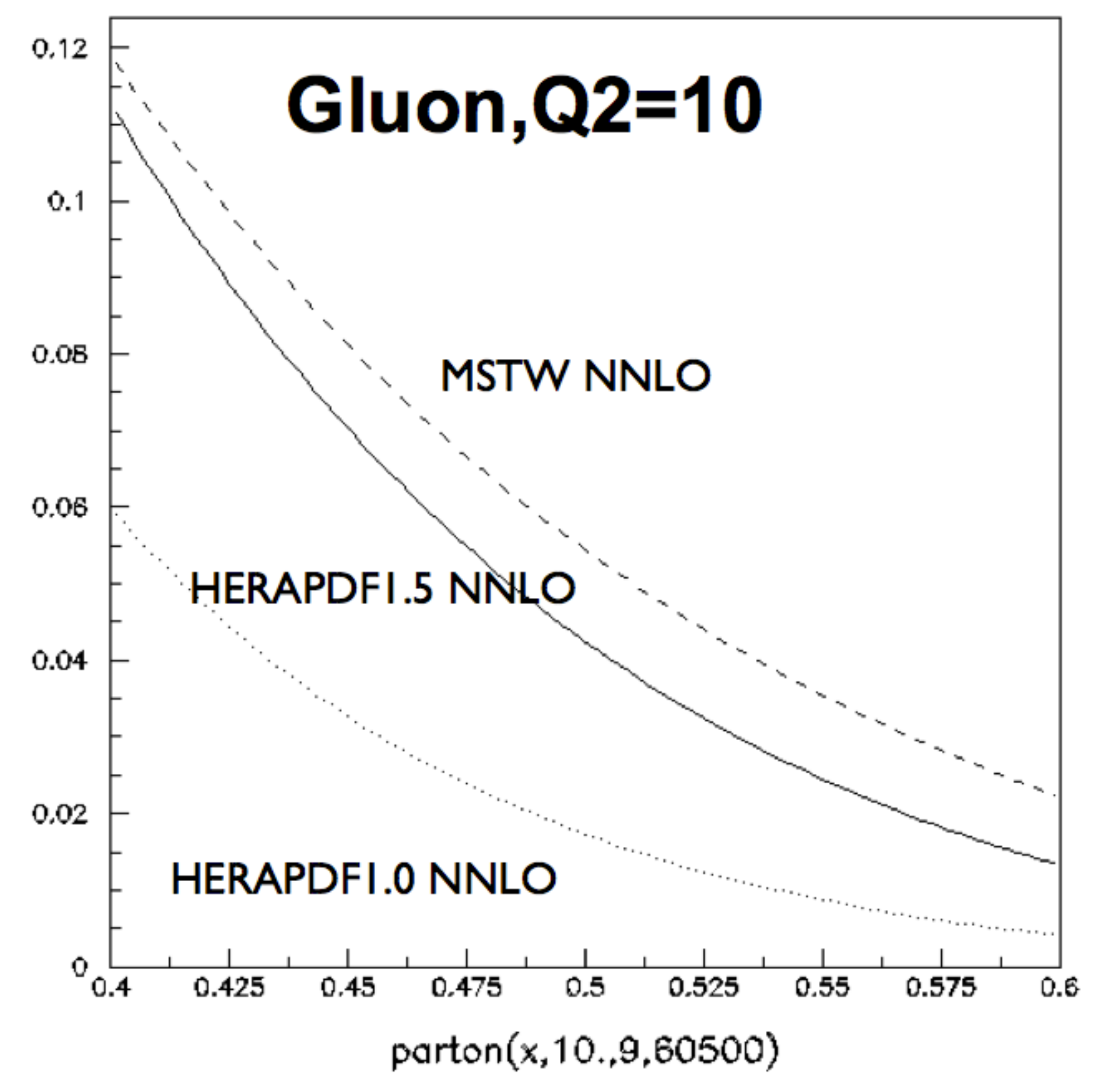}
\includegraphics[width=0.495\textwidth]{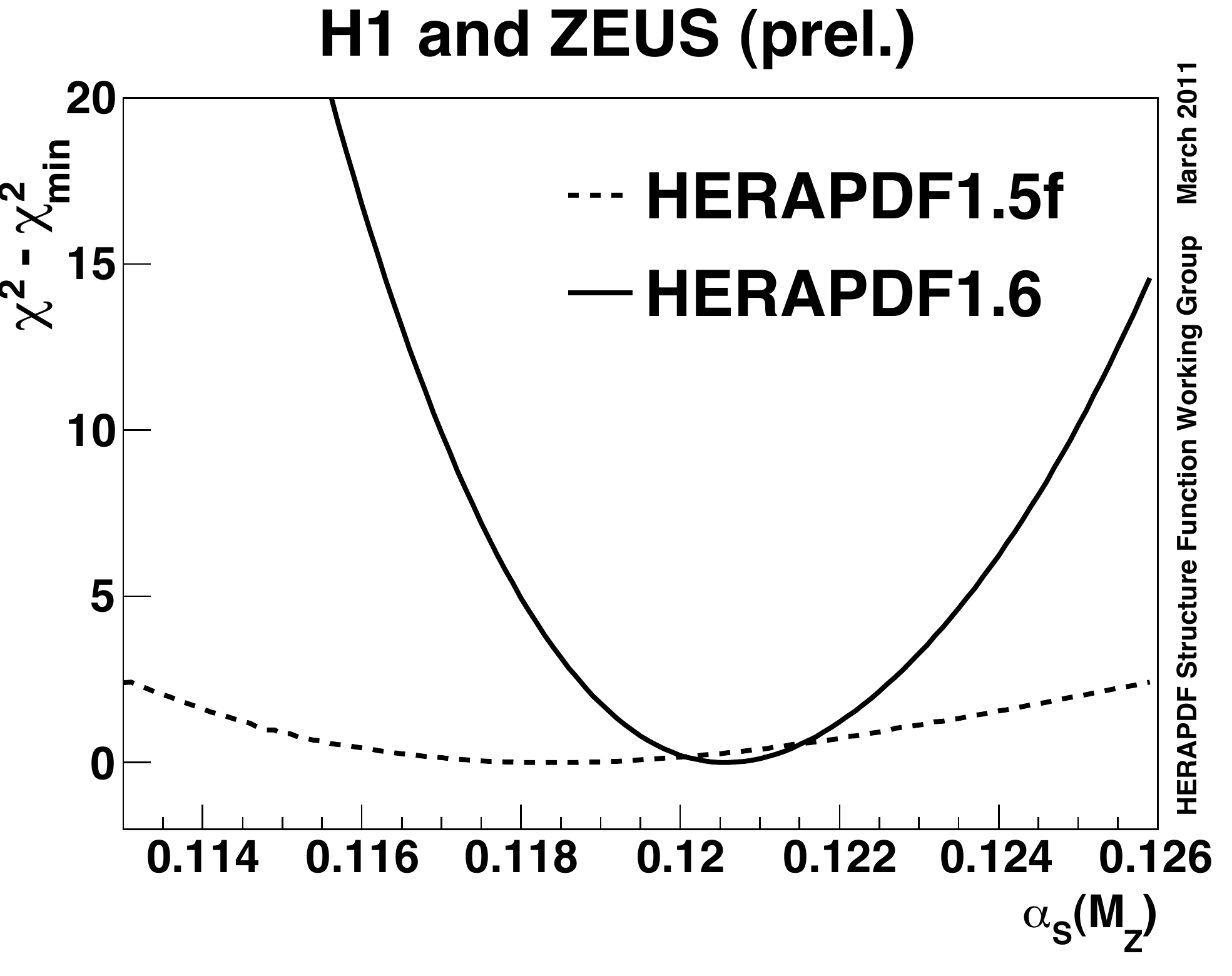}
\caption{On the left hand side: the gluon distribution from the HERAPDF sets (based on HERA I only - HERAPDF1.0, and 
based on HERA I+II,  HERAPDF1.5) to MSTW08 at NNLO in the high $x$ region.
 On the right hand side: the 
 $\Delta\chi^2$ distribution as a function of the value of $\alpha_S(M_Z)$ in the PDF fits for HERAPDF1.5f (dashed line) without jet data  and HERAPDF1.6 (solid line) with jet data. }
 \label{Fig:2}
\end{figure}

The inclusion of the precise high $Q^2$ preliminary HERA II data in the QCD fits results in HERAPDF1.5.
Figure~\ref{Fig:1} shows the comparison between HERAPDF1.5 and HERAPDF1.0 which is based on HERA I data alone.
 The impact is noticeable especially
in the high $x$ region, where the valence contribution dominates. 
In addition, a meticulous study has been performed to 
estimate the uncertainties for the NNLO HERAPDF set,  also shown in Figure~\ref{Fig:1},
where the new NNLO set is compared to the HERAPDF1.0 NNLO.
To answer the question about implications of the new PDF sets on the Higgs exclusion limits from Tevatron,
Figure~\ref{Fig:2} shows a more explicit comparison of the gluon distribution  at high $x$ between HERAPDF and MSTW08 sets at NNLO. The HERAPDF1.5 NNLO set yields a harder gluon at high $x$ compared to the HERAPDF1.0 NNLO set, which is prefered by the 
Tevatron jet data. The differences are mostly due to the use of a more flexible functional form made possible via the availability of a more precise data. 

\begin{figure}[h!]
\centering
\includegraphics[width=0.495\textwidth]{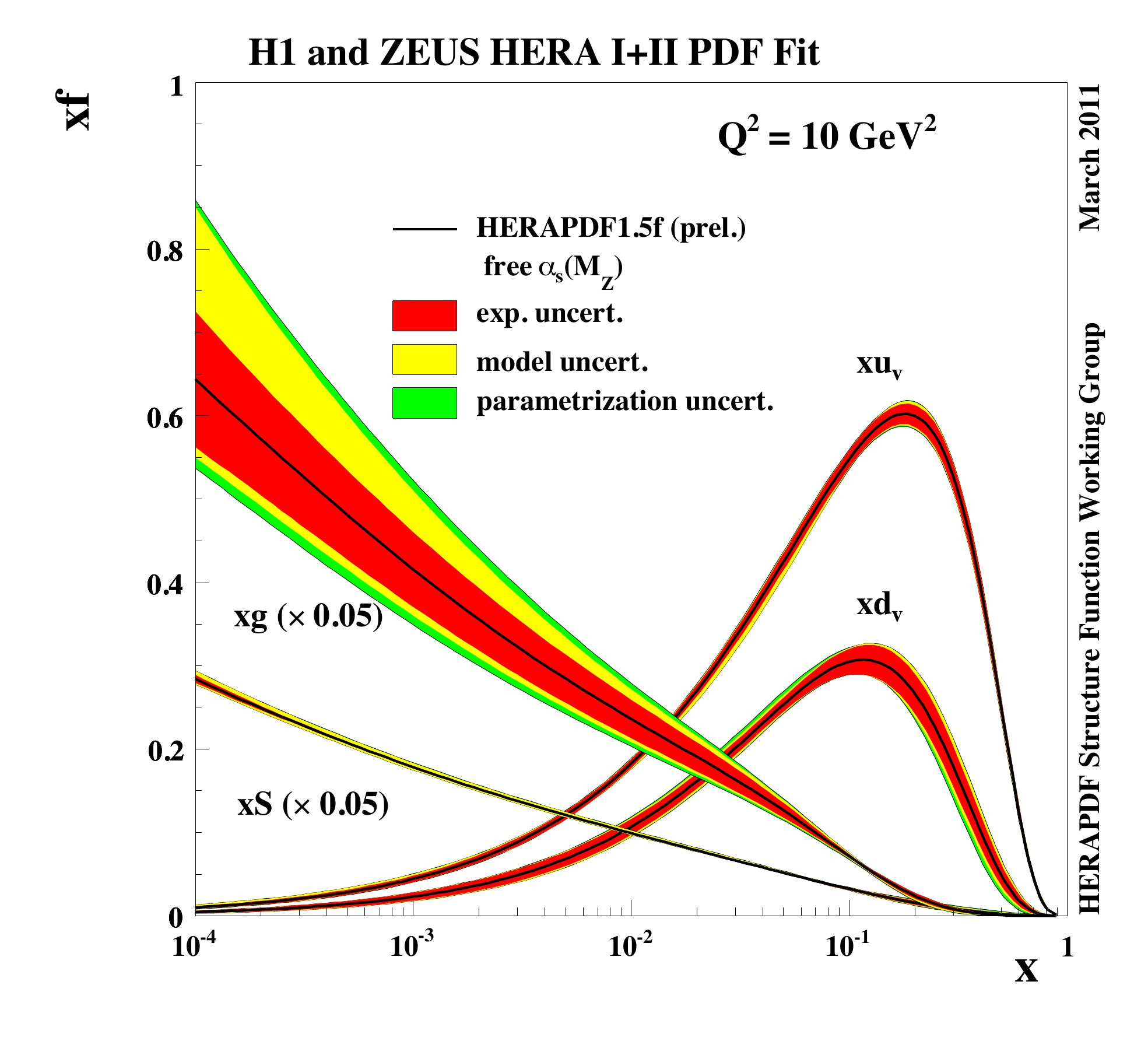}
\includegraphics[width=0.495\textwidth]{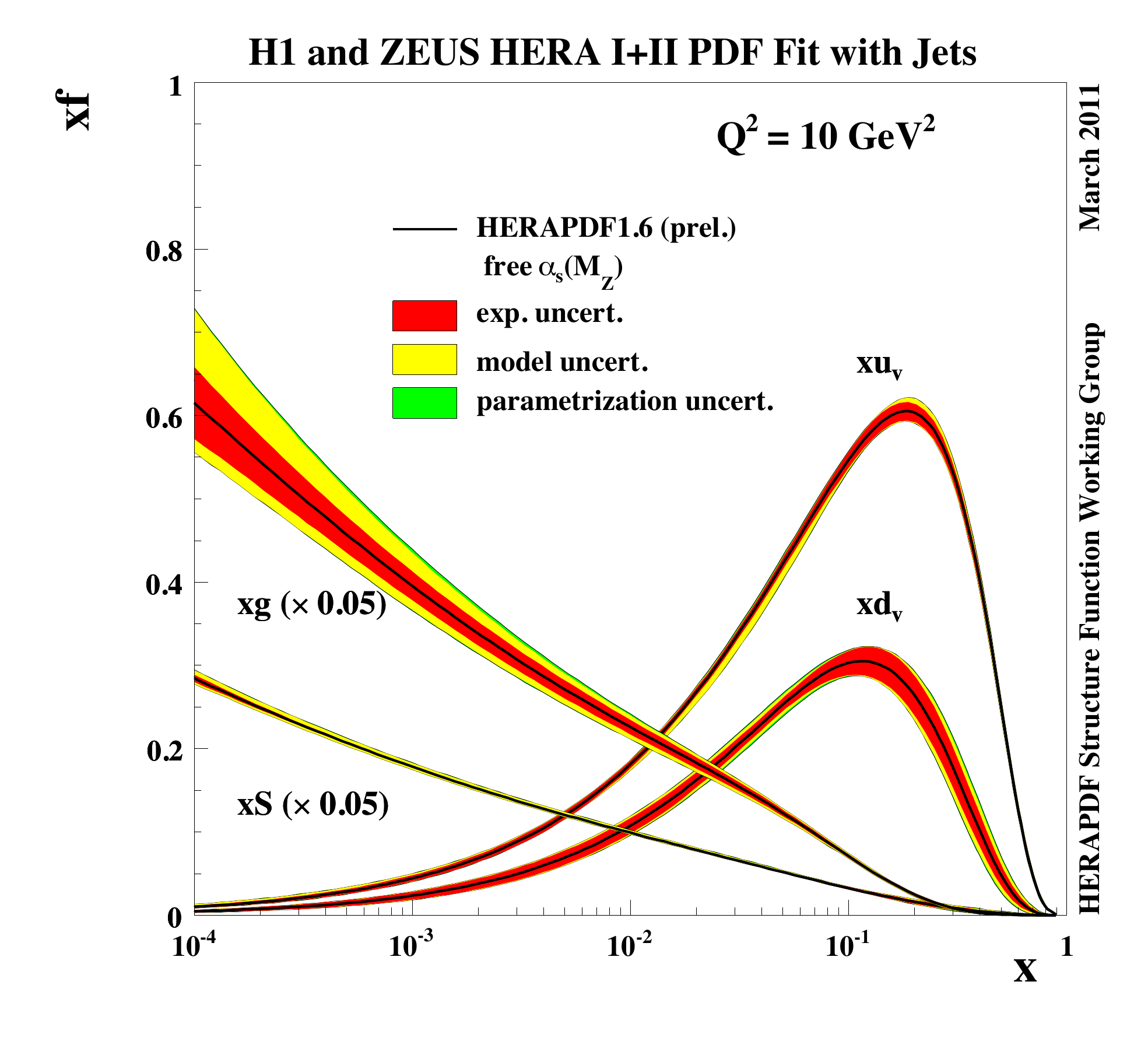}
\caption{Summary plot  for HERAPDF1.5f (left) and HERAPDF1.6 (right) as a function of $x$ for $Q^2 = 10$ GeV$^2$. The strong coupling constant $\alpha_S(M_Z)$ is a free parameter in both fits. The central values of the PDFs (solid lines) are shown together with the experimental, model and parametrisation uncertainties represented by the red, yellow and green shaded bands, respectively.}
 \label{Fig:3}
\end{figure}

Another important result is related to the determination of the strong coupling. New results from HERA based solely on 
the inclusive DIS measurements can not yet pin down the value of $\alpha_S$ as shown in Figure~\ref{Fig:2} (right), where a scan in $\chi^2$ as function of strong coupling is shown.  However, as soon as the jet data are included in the QCD fit, resulting in HERAPDF1.6, it can be seen that  
 the strong coupling  is well constrainted. In fact, the addition of the HERA jet data into the fit allows the strong coupling and gluon to be simultaneously constrained. Figure~\ref{Fig:3} shows the impact of the inclusion of the jet data on the gluon distribution when the fit is performed with the strong coupling as a free parameter. As a result, the HERA data prefer a rather larger value for $\alpha_S(M_Z)=0.1202\pm0.0013($exp$)\pm 0.0007 ($mod$) \pm 0.0012 ($had$)^{+0.0045}_{-0.0036}($th$)$.
 
 In addition, the inclusion of the preliminary HERA charm data provides constraints for the optimal value of the charm mass used in  theory models. It has been observed that QCD fits without charm data have only a small sensitivity to the value of the charm mass.
However, there is a strong preference for a particular $m_c$ once charm data is included. This study has been performed for various schemes, such as those used in the global fit analyses of MSTW08 and CTEQ. The results conclude that each scheme describes the data well at the corresponding best value of the $m_c$. It is interesting to observe that differences in the PDF sets correspond to differences in the charm mass used in different schemes \cite{myslides}.

All HERAPDF sets derived solely from $ep$ measurements are able to give a good prediction of the $Z$ and $W$ at Tevatron from the $p\bar p$ processes, and provide a competitive description to $pp$ processes at the LHC as well.

\vspace*{-0.cm}\section{Summary}
HERA provides accurate determinations of the proton structure and can predict related Standard Model processes.
Additional preliminary combined measurements from HERA II allow the high $x$ region to be better constrained
resulting in a more precise HERA PDF set, HERAPDF1.5 at NLO and NNLO in $\alpha_S$ with  a detailed analysis of the uncertainties. The HERAPDF sets, which are based solely on $ep$ data also describe the Tevatron data well and provide competitive predictions to the LHC processes. Inclusion of the HERA jet data allows for a precise determination of the strong coupling, for which HERA data prefers a rather larger value of $\alpha_S$. In addition, inclusion of the preliminary combined HERA charm data provides constraints on the optimal value for the charm mass used in theory models and it may account for some of the differences among global PDF fits. Therefore, HERAPDF provides a large variety of meticulous studies based on new measurements from HERA for accurate predictions at the LHC \cite{website}.


\section*{References}


\begin{thebibliography}{99}

\bibitem{herapdf} H1 and ZEUS Collaborations, F. Aaron \it et al.\em, JHEP \bf 1001\rm (2010) 109, arXiv:0911.0884. 
\bibitem{jetsh11} H1 Collaboration, F. Aaron \it et al.\em, Eur. Phys. J. \bf C67 \rm (2010) 1, arXiv:0911.5678.
\bibitem{jetsh12} H1 Collaboration, F. Aaron \it et al.\em, Eur. Phys. J.\bf  C65\rm (2010) 363, arXiv:0904.3870.
\bibitem{jetsz1} ZEUS Collaboration, S.Chekanovetal  \it et al.\em, Physics Letters \bf B547 \rm (2002) 164.
\bibitem {jetsz2} ZEUS Collaboration, S.Chekanovetal  \it et al.\em,  Nuclear Physics \bf B765 \rm (2007) 1.
\bibitem{qcdnum} QCDNUM package, M.~Botje, (2010),  arXiv:1005.1481.
\bibitem{tr} R.~S.~Thorne code, revised in 2008.
\bibitem{myslides} V. Radescu slides, \small\verb$http://moriond.in2p3.fr/QCD/2011/MondayMorning/Radescu.pdf$ 
\bibitem{website}The LHAPDF grid files are located at  \\ \small
 \verb$https://www.desy.de/h1zeus/combined$\_\verb$results/index.php?do=proton$\_\verb$structure$ 
 
\end{thebibliography}
\end{document}